\documentclass{lhep}       

\journal{LHEP}%Letters in High Energy Physics}
\vol{xx}
\jyear{2018}
\pages{xxx} 

\received{xx August 2022}
\published{xx March 2018}

\def\be{\begin{equation}}
\def\ee{\end{equation}}
\def\bea{\begin{eqnarray}}
\def\eea{\end{eqnarray}}

\usepackage{amsmath}
\usepackage{amssymb}
\usepackage{xcolor}
\definecolor{dgreen}{rgb}{0,0.6,0.0}

\begin{document}

\title{Capture of dark matter particles by a galaxy in the case of a bimodal distribution of their velocities}

\author{Ruth Durrer\auno{1},  Serge Parnovsky\auno{1,2,3} and Aleksei Parnowski\auno{4}}
\address{$^1$D\'epartement de Physique Th\'eorique and Centre for Astroparticle Physics, Universit\'e de Gen\`eve\\
~~24 quai Ernest-Ansermet,  CH-1211 Gen\`eve 4, Switzerland}

\address{$^2$Taras Shevchenko National University of Kyiv, Astronomical observatory, Observatorna str. 3, Kyiv 04053, Ukraine}
\address{$^3$CERN 1211 Geneva 23. Switzerland}
\address{$^4$Space Research Institute of National Academy of Sciences of Ukraine and State Space Agency of Ukraine, prospekt Akademika Glushkova 40, build. 4/1, 03187 Kyiv, Ukraine}
\begin{abstract}
We have analyzed the rate of capture of dark matter (DM) particles by the galaxy in the case of the existence of two different types of DM or a bimodal velocity distribution function for DM. It is shown that, in addition to the scenario considered in our previous work which is based on the assumption of an unimodal distribution, more complex scenarios are possible in which the transition to the state of intense capture and/or exit from it can occur in two stages.
A detailed description is given of the change in the curve describing the rate of capture of dark matter particles as a function of the rate of increase in the baryon mass of the galaxy for various values of the rate of decrease  of the DM density.
\end{abstract}

\maketitle

\begin{keyword}
Dark matter\sep galaxy formation
\doi{10.2018/LHEP000001}
\end{keyword}

\section{Introduction}
From the analysis of galaxy rotation curves, it is known that
dark matter (DM) is responsible for 85% of the mass of a typical galaxy. It forms a dark halo that surrounds the galaxy and
extends far beyond the limits of its visible part, consisting of
baryonic matter (BM). Evidence for the existence of dark matter is not limited to rotation curves. Reviews [1, 2, 3] describe
other astronomical observations attributed to the influence of
this still enigmatic entity. There are other points of view, in
which the observed effects are explained by the modification
of the laws of physics or with the help of alternative theories of
gravity. There is also a hypothesis that dark matter consists of
a large number of black holes; as a rule, typically these are primordial black holes with masses comparable to or even smaller
than the solar mass.

In this article, we proceed from the assumption that DM
consists of some still unknown particles that are not affected
by strong and electromagnetic interactions. They interact between themselves and with baryonic matter gravitationally. A
weak interaction is possible. However, the so-called mirror DM
incapable of the weak interaction is also acceptable. In principle, it is also possible that DM particles can interact with
each other using forces unknown to us, which do not act or
act very weakly on BM. Candidates for particle dark matter are
described in the paper [3]. The possibility that DM is a mixture
of particles of different types is also acceptable.

These particles fill our Universe. Sometimes, they fly from
intergalactic space into galaxies, and the gravitational field of
the latter affects the speed of dark matter particles, deflecting
them or changing their speed. Consider DM particles flying
in and out of a dark halo. Their velocities inside the halo are
greater than those in the intergalactic space due to falling into a gravitational well. Therefore, the density of DM particles passing through the halo is less than that outside the halo. If we consider the galaxy as a source of a static gravitational field, then the incoming particle has enough energy to leave the galaxy.

Galaxies were formed after recombination as a result of the
growth of density fluctuations. So, a natural question arises
as to where this dark matter came from. The answer clearly
refers to the time when the gravitational field of a galaxy is
not constant. For example, it can be associated with the stage
of galaxy formation. Galaxies formed after recombination as a
result of the growth of small initial density fluctuations. This
idea is in good agreement with the small fluctuations observed
in the temperature of the cosmic microwave background [4, 5].
An overdensity leads to the infall of the surrounding matter.
In this case, BM is basically captured, increasing the mass of
the overdensity, while the DM can fly through it without being
captured.

As long as perturbations are small, they can be studied with
linear or higher-order perturbation theory. But for the formation of galaxies, such an approach is not suitable for two reasons. First of all, the overdensities of galaxies, $\rho_{\rm gal}/\rho_m \gtrsim 10^5$, are much larger than 1 so that the perturbation theory cannot be expected to converge. Furthermore, we assume DM particles to be collisionless. In this case, they should be described by Vlasov’s equation in phase space. While this leads to a similar Jeans scale as the fluid approach, just replacing the sound speed with the velocity dispersion [6, 7], nonlinear aspects are very different as soon a shell crossing becomes relevant which leads to singularities in the fluid approach.

Therefore, the nonlinear regime of the cosmological structure formation is usually treated either via N-body simulations, which have, at present, achieved an impressive amount of detail [8, 9, 10, 11] or by simple analytical models, usually spherical collapse or secondary infall models [12, 13, 14, 15, 16].

In article [17], we have proposed to use a new approach different from those mentioned above. Unlike the N-body simulations, we consider a continuous density distribution of particles without their concentration in massive objects, which allows one to use analytical treatment of the motion of DM particles and guarantees the absence of divergent acceleration. We do not consider the complex processes of accretion and thermalization of BM, concentrating on the investigation of the possibility of capturing DM particles. The influence of baryonic matter is reduced to a single parameter, namely, the rate of increase of the overdensity due to BM accretion. As a result, we show the possibility of a catastrophic increase of the DM particle capture rate associated with the strong nonlinearity of this process. 

The main mechanism of the increase in the DM mass is the capture of DM particles flying into the halo from intergalactic space. However, without BM accretion by the galaxy, they would fly through the galaxy without being captured. If the baryonic mass of the galaxy increases due to the accretion of BM or the merger of galaxies (see the details of galaxy formation in [7]), then some of the passing DM particles cannot escape due to a lack of kinetic energy. In this case, the depth of the potential well has time to increase during the time the particle passes through the galaxy. Correspondingly, the height of the potential barrier, caused by the gravitational interaction of the particle with the galaxy, also increases.

This is determined by the total mass of the galaxy $M$, which can increase due to the accretion of BM and the capture of DM. Accordingly, the rate of its increase $\dot M$ is equal to the sum of the rates of increase in DM, designated as $\dot M_{DM}$, and the rate of increase in BM, designated as $\dot M_b$:
\be\label{e:Mdotsum}
\dot M=\dot M_{DM}+\dot M_b \,.
\ee	
We consider $\dot M_b$ as an external quantity that can change with time, and we try to calculate $\dot M$ as function of $\dot M_b$.We assume that the mass of the galaxy $M$ is known at the moment of time under consideration (assuming that it does not change significantly during the passage of the DM particle, which cannot be captured due to its rather high initial velocity $v_0$ far away from the galaxy). We also know the halo radius R, which we consider to be constant, and the parameters of dark matter particles in intergalactic space. 

They include the particle mass $m$, particle density $N$, and
the particle velocity distribution. The last two quantities are determined by the distribution function $f(v_0)$.  Let the number of particles with velocities in the range from $v_0$ to $v_0+dv_0$ in a unit volume in extragalactic space is equal to $dN=f(v_0 )dv_0$. So, the total density of DM particles is $N=\int_0^\infty f(v_0)dv_0$. 

We consider DM particles to be nonrelativistic, and we use the formulas of classical mechanics. Indeed, astronomical observations show that the characteristic velocities of dark matter are much less than the speed of light. This type of dark matter is called cold dark matter, in contrast to hot dark matter with particles moving at relativistic speed. Estimates of the minimum speed of a DM particle in intergalactic space, which allows it to fly through the Milky Way without capture, were made in [17]
and give a value not exceeding 100 km/s, which is significantly
less than the speed of light. Faster particles fly through the
Galaxy and do not contribute to the increase in the DM mass
in it. Actually, nothing prevents us from taking into account the
effects of Special Relativity and considering the case of hot DM,
but this apparently will not lead to a qualitative change in the
main result of our model, specifically the catastrophic increase
of the DM particle capture rate. We assume that the rest mass
of the DM particle is large enough not to take into account its
wave properties but otherwise arbitrary.

We take into account the Hubble expansion to account for
changes in the DM density in intergalactic space. $N$  decreases
rapidly with time because of both Hubble expansion and the capture of a part of the DM particles by galaxies. The first of
these effects is stronger and ensures a dependence according to
the law $N\propto (1+z)^3$.  The second makes the decrease in N even faster. Other parameters of the galaxy like its mass, M, or of the velocity distribution of DM, $f(v_0)/N$ change much more slowly with time.

The model described in the article [17] assumes that the
halo is spherically symmetrical and homogeneous; i.e., has no
inhomogeneities, on which the DM particles could perform a
gravitational maneuver. Therefore, particles moving radially
can either be captured at $v_0<v_{pc}$ or fly through the galaxy at $v_0>v_{pc}$ (the subscript pc indicates passage through the center). The estimate detailed in [17] gives
\be
v_{pc} \simeq a\sqrt{\dot M}, \quad a=\left(\frac{2GR}{M}\right)^{1/4}\,. \label{e:vpc}
\ee
A non-linear equation
\be
\dot M_{DM} \simeq  \pi R^2\int_0^{v_{pc}}\left(1-\frac{v_0^4}{v_{pc}^4} \right)  mf(v_0 ) v_0 dv_0 \,	
\label{e:Mdmdot}		
\ee	
was also obtained in the article [17]. From the analysis of  eqs. (\ref{e:Mdotsum}-\ref{e:Mdmdot}) we obtained a qualitative picture of the process of capture of DM particles and of the accumulation of DM mass in the galaxy. In this article, we will continue the analysis, considering additionally some more exotic variants.

These options arise due to the fact that the equation \eqref{e:Mdmdot} includes an unknown distribution function $f(v_0)$. We have no observational data to establish its type or convincing theories to predict it. We do not know if DM particles are in a state of thermal equilibrium. It is only known that dark matter has characteristic velocities much smaller than the speed of light and belongs to the so-called cold or warm dark matter. The latter has properties intermediate between those of hot dark matter and cold dark matter, but the differences are mainly related to the details of structure formation. 

Nevertheless, one can make a number of reasonable assumptions about the form of $f(v_0)$. We suppose that the function $f(v_0)\geq 0$ is continuous, vanishes at $v_0=0$, reaches a maximum at some value $v_0=v_{\max}$, and quickly decreases to zero at high-speed end, most likely exponentially in $v_0^2$. At small $v_0$ we can put 
\be
f(v_0)\approx a_1 v_0^{2} \label{a1} 
\ee
due to the three independent Cartesian velocity components. The coefficient $a_1 \propto N$.

This is enough to analyze the solution of the system (\ref{e:Mdotsum}-\ref{e:Mdmdot}) and derive a number of qualitative conclusions. They are based primarily on the shape of  the curve $\dot M(\dot M_b)$. We immediately note that this form is constantly changing due to the time dependence of the parameters included in  equations (\ref{e:vpc}) and (\ref{e:Mdmdot}), primarily $N$.

In this case, as we have found in [17] a qualitative change
in the shape of the curve occurs. When $N$ is less than some critical value $N_1$, the curve increases monotonically. In this case, the capture of dark matter occurs weakly and only at $\dot M_b>0$. At $N>N_1$ the curve acquires a characteristic s-shape. In this case it consists of three branches. Two of them have a positive derivative and correspond to stable solutions. They are connected by a section with a negative derivative corresponding to unstable solutions. The behavior of the solution in this case is well studied in catastrophe theory. It is well described in many books including [18, 19]. This shape is called the fold catastrophe 

The solution $\dot M_b=\dot M=0$ is stable, as is any point on the stable branches. Therefore, in the absence of accretion, there is no capture of DM particles on the lower branch. But for $\dot M_b>0$ we have $\dot M_{DM}>0$. So capture in this case is a consequence of synergy, i.e. combined action of BM accretion and passage of DM particles. If the system is on the lower stable branch and an external influence, i.e. the value of $\dot M_b$ increases so much that it reaches the right boundary of the lower branch (we will refer to it as point B) and continues to increase, then the state of the system jumps to the upper stable branch and remains on it even after decreasing $\dot M_b$. 

The upper branch describes the galaxy in a state of intense
capture of DM particles. In this case, the ratio between the
growth rates of the dark and baryon masses can be very large.
This explains the high abundance of dark matter in galaxies. In
our opinion, the galaxies captured almost all of their DM in the
state of intense capture. In [17], the mass of dark matter that
could be captured by the Galaxy was estimated if the process
of intense capture began at the redshift parameter z0. It turned
out that the proposed mechanism can ensure the capture of all
dark matter in the galaxy at $z_0=15...25$. The process of intense capture could have begun even before the appearance of the first stars and reionization.

The upper stable branch has a left boundary, i.e. the minimum value $\dot M_b$. We will refer to it as the point C. As $\dot M_b$ decreases, the system located on the upper branch can reach point C and, as $\dot M_b$ decreases further, jump to the lower branch. In this case, the galaxy leaves the stage of intense capture. Note that this is possible only if $\dot M_b>0$ in point C. As $N$ decreases, the curve changes so that point C shifts to the right, crossing at some moment of time the y-axis. Only after this crossing is it possible to end the stage of intensive capture with a decrease in $\dot M_b$.

Something like a phase transition occurs at $N=N_1$ with  a change in the regime of particle capture. But by this time, the intensive capture mode will most likely have had time to end.

\begin{figure}\begin{center}
\includegraphics[width=8cm]{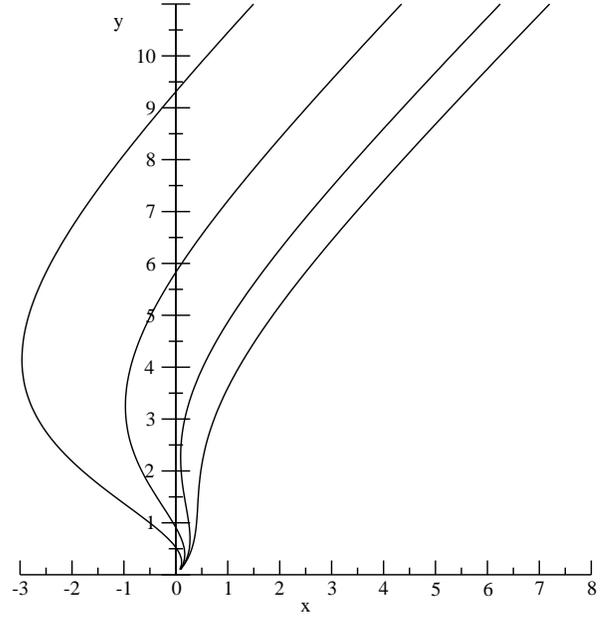}
\end{center}
\caption{\label{f:1}Dependence of $x$ and $y$ defined in \eqref{xy} proportional to the rates of increase in the total and baryonic masses of the galaxy for the case of the Maxwell distribution of DM particle velocities. The curves from left to right correspond to the different values of the parameter $\mu$ from \eqref{mu} equal to 10, 7, 5 and 4.}
\end{figure}

We can illustrate all of the above with the example of the Maxwell-Boltzmann distribution 
\be
f(v)=\frac{4Nv^2}{\sqrt\pi v_{max}^{3}}\exp{\left[-\left(\frac{v}{v_{max}}\right)^2\right]}, \,v_{max}=\sqrt{\frac{2k\Theta}{m}}\label{mb}
\ee
with temperature $\Theta$ and maximum at $v=v_{max}$. Fig. \ref{f:1} shows the dependence of quantities proportional to $\dot M$ and $\dot M_b$. 
\be
x=\gamma \dot M_b,\quad y=\gamma \dot M,\quad\gamma=\sqrt{\frac{2GR}{M}}v_{\max}^{-2}. \label{xy}
\ee
All parameters included in (\ref{e:Mdotsum}-\ref{mb}) are reduced to a single parameter
\be
\mu =\frac{2R^2mN}{v_{\max}}\sqrt{\frac{2\pi GR}{M}}, \label{mu} 
\ee
which decreases due to the decrease of $N$.  In Figure 1, one can
see both the transition from the s-shaped curve to a monotone
one and the intersection of point C with the y-axis. We see that
after the jump to the upper stable branch, the rate of increase
in the mass of the galaxy due to the capture of DM particles
increases greatly at the same rate of increase in the baryonic mass. For the curves depicted in Figure 1, it increases 11 times
at $\mu=7$, and 85 times at $\mu=10$. This shows how significantly the catastrophic transition to the intensive capture regime affects the ratio of the growth rates of
dark and baryonic matter. 

Again, we are using the Maxwell-Boltzmann distribution for illustration purposes only. It has all the properties that, according to our assumption, are inherent in the function $f(v)$, including the asymptotic behavior of (\ref{a1}) at low speeds and fast exponential decay at large values of $v$. However, we do not claim that it describes the actual velocity distribution of dark matter particles or that dark matter particles are in a state of thermal equilibrium.

We can describe the process of accumulation of dark matter inside the halo of the galaxy. As a result of the growth of small density fluctuations, regions of increased density emerge, in which galaxies can form. The surrounding matter, both baryonic and dark, begins to fall into them. In the absence of accretion of baryonic matter, DM particles fly through protogalaxies and are not captured.

But the situation changes with an increase in the mass of the baryonic component of the galaxy. A fraction of the DM particles passing into the galaxy at sufficiently low speed is being captured. The mass of dark matter inside the galaxy begins to grow, but the rate of its increase is smaller than the rate of increase in the mass of BM. 

If the rate of increase in the mass of the BM exceeds a certain threshold value, corresponding to point B, a jump on the upper branch occurs. The galaxy goes into the regime of intense capture of DM particles. The system remains on the upper branch until either the particle density drops below $N_1$, or until the point C crosses the y-axis, and the baryon mass growth rate becomes less than the abscissa of the point C. The galaxy comes out of the regime of intense capture of DM particles.

\section{Exotic examples of DM capture}
%associated with unusual particle velocity distributions}
\begin{figure}\begin{center}
\includegraphics[width=8cm]{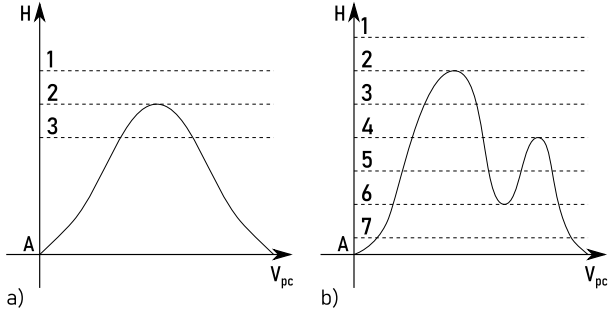}
\end{center}
\caption{\label{f:2}Unimodal (a) and bimodal (b) functions $H$ associated with the velocity distribution of dark matter particles.}
\end{figure}
Let us try to add something to this  picture and show that sometimes the process of transition to the stage of intense capture or exit from it occurs in a more complicated way. This possibility arises in some cases when the distribution function $f(v)$ is bimodal. Can such a function be realized? As already mentioned, we do not know the form of $f(v)$. It is possible that some processes in the early Universe could lead to a such distribution of DM particle velocities. 

But there is also another possibility. There may be two (or more) different types of dark matter with different particles and different velocity distributions, then we can include both in equation \eqref{e:Mdmdot}. Both types of particles have the properties that
we attributed to DM particles. They gravitationally attract BM
and dark matter of any kind and do not participate in strong
and electromagnetic interactions. The capture of DM particles
of any kind increases the mass of dark matter in the galaxy and
its total mass but does not affect the rate of increase in the mass
of the baryonic component

In this case, it is easier to consider the velocity distribution
not of the number of particles but of their mass density $mf(v)$. Since only the product of $f(v)$ and the particle mass $m$ enters into the formulas, the transition is simple. It is obtained by summing the distribution for each particle type. If the properties of different DM particles are very different, then the velocities corresponding to the modes of the two distributions are also likely to be different. By summing two distributions with different modes, we can get a total bimodal distribution. 

If DM particles of both types do not born and do not decay, then the ratio between their total number in the Universe (but not in the halo of an individual galaxy) is preserved and the average DM mass density is proportional to their average concentration N.
 So, it is worth analyzing the behavior of the curves describing the dependence of $\dot M$ on $\dot M_b$ for the case of an arbitrary distribution $f(v)$ that satisfies the formulated conditions.

Note that all points on the curve that are of most interest to us, like points B and C, have vertical tangents. The same applies to the moment of the phase transition at $N=N_1$, when a point with a vertical tangent appears on the curve. At these points we have the condition $d\dot M/d\dot M_b=\infty$ or $d\dot M_b/d\dot M=0$. From \eqref{e:Mdotsum} we then get $d\dot M_{DM}/d\dot M=d\dot M_{DM}/dv_{pc}\cdot dv_{pc}/d\dot M=1$. From \eqref{e:Mdmdot} and \eqref{e:vpc} we then obtain 
\be
H(v_{pc})=\frac{1}{2\pi R^2mN}\left(\frac{M}{2GR}\right)^{1/2}\,,\label{H1}
\ee
where we have introduced the function $H$ defined by
\be
H(v_{pc})\equiv N^{-1}v_{pc}^{-6}\int_0^{v_{pc}}v_0^5f(v_0)dv_0.\label{H2}
\ee
The factor $N^{-1}$ is introduced such that this function does not directly depend on $N$. It compensates for the fact that $f\propto N$. As a result, the function $H$ changes with time, but slowly, not as fast as the particle density $N$.
\begin{figure}\begin{center}
\includegraphics[width=8cm]{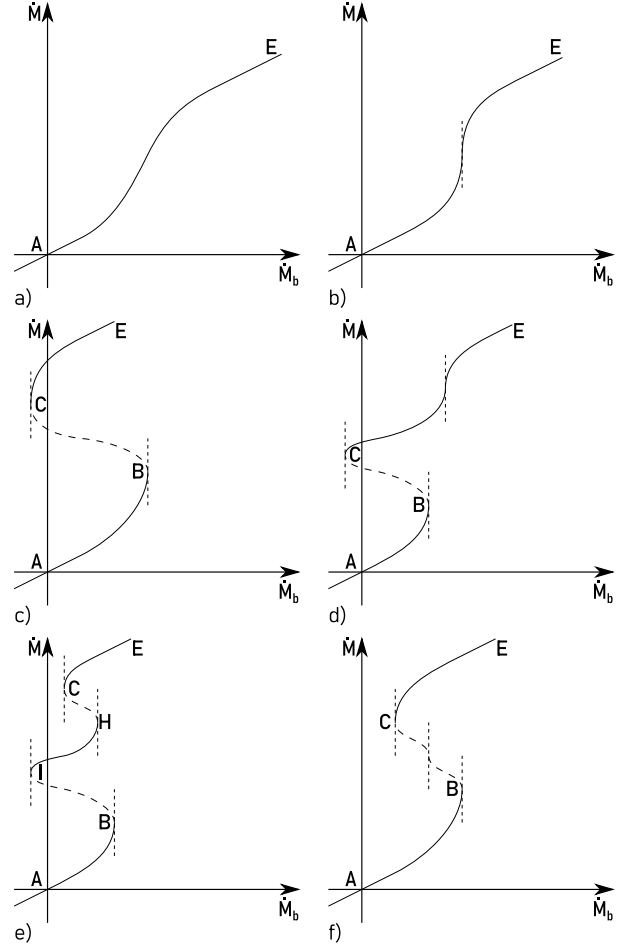}
\end{center}
\caption{\label{f:3}Characteristic curves of dependence $\dot M$ on $\dot M_b$ at different values of particle density $N$. a) $N<N_1$, line 1 on Fig. \ref{f:2}. b) $N=N_1$, line 2 on Fig. \ref{f:2}. c) S-shaped curve for lines 3 and 7. d) The same as c) but with a vertical tangent point outside BC segment for line 4.  e) Double s-shaped curve for line 5. f) The same as c) but with a vertical tangent point inside BC segment for line 6.}
\end{figure}

Let us consider its properties. The function $H(v_{pc})\ge 0$ and tends to zero as $v_{pc}\to 0$ because of asymptotic behavior \eqref{a1}. It also tends to zero
as $v_{pc}\to \infty$ because the integral in \eqref{H2} tends to a finite value, but the factor $v_{pc}^{-6}$ ensures that it falls according to the law $H(v_{pc})\xrightarrow[v_{pc}\to \infty]{}O(v_{pc}^{-6})$. This means that it has a maximum at a certain $v_{pc}=v_1$.

Can it have several local maxima with minima in between? At each extremum, the condition  
\be
\frac{dH(v_{pc})}{dv_{pc}}= N^{-1}v_{pc}^{-7}\left( v_{pc}^6f(v_{pc})-6\int_0^{v_{pc}}v_0^5f(v_0)dv_0\right)=0\label{H3}
\ee
must be satisfied. After integration by parts, it reduces to
\be
\int_0^{v_{pc}}v_0^6 {f^\prime(v_0)}dv_0=0,\label{H4}
\ee
where the prime means the derivative with respect to the argument. If the function $f(v_0)$ is bimodal and has two maxima at the points $v_{max1}$ and $v_{max2}>v_{max1}$ and a local minimum at $v_{min1}$ between them, then ${f^\prime}$ is positive on the segments $(0,v_{max1})$, $(v_{min1},v_{max2})$ and negative on the segments $(v_{max1},v_{min1})$, $(v_{max2},\infty)$. Condition \eqref{H4} can be met at some point between $v_{max1}$ and $v_{min1}$ only if
\be
\int_0^{v_{max1}}v_0^6 {f^\prime(v_0)}dv_0<\int_{v_{max1}}^{v_{min1}}v_0^6 {|f^\prime(v_0)|}dv_0 \label{H5}.
\ee
If this condition is not satisfied, then there are no extrema of the function $H$ also on the segment $(v_{min1},v_{max2})$. A simple analysis shows that $H$ can be bimodal only if $f$ is bimodal or multimodal, and the maxima of $f$ are at a sufficiently large distance from each other. But this variant is possible in principle. However, the bimodality of $f$ does not guarantee that $H$ is bimodal.
\begin{figure}\begin{center}
\includegraphics[width=8cm]{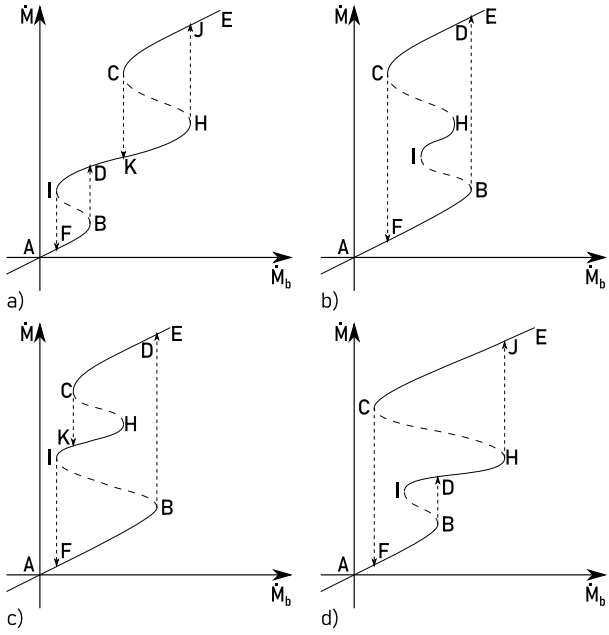}
\end{center}
\caption{\label{f:4}Different options of the curve in Fig. \ref{f:3}e provide a transition to the upper and lower branches with a different number of jumps.}
\end{figure}

Consider the function $H$ at some moment of time. It can be unimodal as in Fig. \ref{f:2}a or bimodal. On Fig. \ref{f:2}b shows an example of such a function. Of course, any of the two maxima can be the global maximum at $v_{pc}=v_1$. In Fig. \ref{f:2} it is the left one, which has a smaller value of $v_{pc}$ and hence $\dot M$. But it is easy to construct curves for the case when the right maximum is higher.

According to \eqref{H1}, the points with vertical tangents on the dependence of $\dot M$ on $\dot M_b$ correspond to the intersections of the curve $H(v_{pc})$ with the horizontal line of height given by the right hand side of the equation \eqref{H1} which decreases with increasing $N$. We denote
\be
N_1=\frac{1}{2\pi R^2mH(v_1)}\left(\frac{M}{2GR}\right)^{1/2}.\label{N1}
\ee
At $N<N_1$ we have no solution and the points with vertical tangents do not exist. On Fig. \ref{f:2} this case corresponds to the horizontal line 1 passing above the curve. The characteristic form of dependence $\dot M$ on $\dot M_b$ for this case is shown in Fig. \ref{f:3}a.

At $N=N_1$ the line 2 touches the curve at its maximum. This value  corresponds to the transition to an s-shaped curve from the monotonic one. The characteristic form of the curve for this case is shown in Fig. \ref{f:3}b. At $N>N_1$ we have several possibilities. For the line 3 below the maximum (but above the second mode for the bimodal $H$) we find two solutions of the equation \eqref{H1}. The solution with smaller $v_{pc}$ corresponds to point B, one with larger $v_{pc}$ to point C. For small $v_{pc}$ we can use the approximation \eqref{a1} and obtain the estimate for coordinates of point B. The characteristic form of curve for this case is shown in Fig. \ref{f:3}c.

The following lines are only of interest in the case of the bimodal distribution shown in Fig. \ref{f:2}b. 
Line 4 touches the top of the lower mode. The characteristic curve for this case is depicted in Fig. \ref{f:3}d. Line 5 has four points of 
intersection with $H(v_{pc})$. The characteristic curve depicted in Fig. \ref{f:3}e, its shape 
resembles two letters s one above the other. Line 6 passing through the local minimum provides the 
characteristic curve depicted in Fig. \ref{f:3}f. The characteristic curve for line 7 below the local 
minimum in Fig. \ref{f:2}b qualitatively the same as 
in Fig. \ref{f:3}c. 
 
 Recall that lines 1-7 correspond to an increase in the particle density $N$, which corresponds to earlier stages of the evolution of the galaxy. In the process of this evolution, the curve $\dot M$ on $\dot M_b$ changes its form, successively acquiring the forms shown in Figures \ref{f:3}c, \ref{f:3}f, \ref{f:3}e, \ref{f:3}d, \ref{f:3}c, \ref{f:3}b, and \ref{f:3}a.
 
 Of all these stages, the most exotic is by far the one whose graph is shown in Fig. \ref{f:3}e. However, this conditional image corresponds to 4 possible variants of the mutual arrangement of the abscissas of points with vertical tangents. All of them are shown in Fig. \ref{f:4}. The curve has three stable branches AB, IH, CE and two unstable ones BI, HC.
 
 Let us consider  the different possibilities one by one. The curve in Fig. \ref{f:4}a corresponds to the case when the transition to and exit from the stage of intense capture of DM particles occurs in two jumps BD + HJ and CK + IF. In Figure \ref{f:4}b, both transitions occur in one jump. The states on the branch IH are stable, but unattainable for the system. In Fig. \ref{f:4}c the upward jump BD occurs in one stage, downward in two, Fig. \ref{f:4}d depicts the opposite case. 
 
 Let us discuss the question of which direction the dependence curve shifts during galaxy formation, when the value of $N$ rapidly decays. Consider an arbitrary point on this curve that has some fixed ordinate, i.e. constant value $\dot M$. From \eqref{e:vpc} we find the dependence of $v_{pc}$. The value of $a$ does not depend on $N$. It slowly decreases due to the increase in the mass of the galaxy $M$, but this can be neglected in comparison with the effect of the rapidly falling particle density $N$. The value of $\dot M_{DM}\propto N$ also decreases rapidly. The abscissa of the point $\dot M_b$ shifts to the right with decreasing $N$ according to \eqref{e:Mdotsum}. Therefore, in the process of evolution, the curve $\dot M$ vs. $\dot M_b$ shifts to the right. Fig. \ref{f:1} illustrates this.
 
  This also applies to the abscissas of the points B, C and H, I, if the last two are present. But their ordinate also changes. Fig. \ref{f:2} shows that the value of $v_{pc}$ and hence also $\dot M$ increases for points B and H and decreases for points C and I as $N$ decreases.
 
Is the number of stages required to jump to the upper stable branch and to descend to the lower stable branch predetermined by the initial conditions? A jump to another branch requires the s fulfillment of two conditions: the presence of a point with a vertical tangent at some value $\dot M_{b0}$ on the curve $\dot M(\dot M_b)$ and a change in $\dot M_b$ so that its value coincides with or passes through the value of $\dot M_{b0}$. Both the shape of the curve $\dot M(\dot M_b)$ and the external parameter $\dot M_b$ change with time. If we know exactly how they change, we can determine the number of jumps during the evolution of the galaxy. 

However, the rate of increase in the mass of the baryonic
component $\dot M_{b0}$ depends on the rate of accretion and merger of
galaxies, and the shape of the curve is also determined by the
parameters of the galaxy that change depending on accretion,
for example, its total mass and size. Therefore, the number of jumps during the formation of a galaxy depends on the values determined by the environment of the given galaxy, especially $\dot M_b$, and is not determined only by the initial conditions. One can expect one jump on a branch with an intense capture and a second jump when returning to a branch with a moderate capture in the case of a unimodal distribution without strong changes in the $\dot M_b$ value. But more than two jumps  are also possible in the case of a bimodal distribution of DM particles velocities. The transition to any of the above-mentioned branches can occur in two stages with a jump to an intermediate stable branch.
 
 \section{Conclusions}
 We have analyzed the rate of capture of DM particles by the galaxy for two different types of DM particles or a bimodal velocity distribution function of DM particles. It is shown that, in addition to the scenario considered in our previous work which was based on the assumption of unimodality of the velocity distribution, more complex scenarios are possible in which the transition to the state of intense capture and/or exit from it occurs in two stages.

A detailed description is given of the change in the curve describing the rate of capture of dark matter particles as a function of the rate of increase in the baryon mass of the galaxy for various values of the density of dark matter which decreases with time, $N$. As $N$ decreases, the curve shifts to the right, increasing the value of $\dot M_b$ at a practically constant $\dot M$.

\section{Acknowledgement}
We are grateful to the unknown reviewers for their comments
and questions, which made it possible to improve the presentation of this article. R. Durrer acknowledges support from
the Swiss National Science Foundation. S. Parnovsky expresses
gratitude to the people and the government of the Swiss Confederation for supporting Ukrainian scientists in wartime. He
thanks SwissMAP for funding his visit to the University of
Geneva in Spring 2022, and the Departement de Physique ´
Theorique and CERN for the opportunity to prolong this visit.

 \section{CONFLICTS OF INTEREST}
The  authors  declare  that  there  are  no  conflicts  of  interest  regarding the publication of this paper.

\bibliographystyle{unsrt}

\end{document}